\documentclass[pra,eqsecnum,twocolumn]{revtex4}

\usepackage{graphicx}
\usepackage{upgreek}
\usepackage{amsmath}

\newcommand{\AM}{\mathrm{AM}}
\newcommand{\be}{\begin{equation}}
\newcommand{\ee}{\end{equation}}

\newcommand{\tE}{\tilde{\mathbf{E}}}

\begin{document}

\title{Transverse radiation force  in a
tailored  optical fiber }

\date{\today}

\author{Iver Brevik} %\email{iver.h.brevik@ntnu.no}
\author{Simen {\AA}. Ellingsen}
%\email{simen.a.ellingsen@ntnu.no}
\affiliation{Department of Energy and Process Engineering,
Norwegian University of Science and Technology, N-7491 Trondheim,
Norway}

\begin{abstract}
We show, by means of simple model calculations, how a weak laser
beam sent though an optical fiber exerts a transverse radiation
force  if there is an azimuthal asymmetry present in the fiber
such that one side  has a  slightly different refractive index
than the other. The refractive index $\Delta n$ needs only to be
of very small, of order $10^{-3}$, in order to produce an
appreciable transverse displacement of order $10~\mu$m. We argue that the effect has probably already been seen in a recent
experiment of She {\it et al.} [Phys.\ Rev.\ Lett.\ {\bf 101},
243601 (2008)], and we discuss correspondence between these observations and the theory presented. The effect could be  used to bend optical fibers
in a predictable and controlled manner and we propose that it
could be  useful for micron-scale devices.
\end{abstract}

%\end{center}

\maketitle

\section{Introduction}

Recent years have seen an increased interest in radiation forces
in optics. Optical tweezers, atom traps, optical manipulation of
soft materials such as interfaces between liquids - especially
near the critical point where surface tension is small - are
typical examples. This trend will in all probability continue
in the time to come.

Our objective in the present Communication is
to point towards the possibility  - to our knowledge not so far
contemplated in the literature  - to create a tailored transverse
optical force in a fiber transmitted by a laser beam. The beam may
be pulsed, or it may be continuous.
 The clue of the principle is to introduce an
accurate mechanical imbalance in the fiber, implying a slight
asymmetry in the refractive index $n$. (In practice, such a
deviation from axisymmetry may easily result inadvertently, during
the mechanical drawing of the fiber.) If one side of the fiber is
harder than the other, there may be a slight refractive index
difference $\Delta n$ between the two sides, resulting in a
transverse optical force. As fibers of micron scale cross sections
are very light and bend easily, a sideways motion may easily
occur. The effect, besides being of basic interest, may be of
practical utility.  We will  describe the effect, making use of
simple models for the fiber, and thereafter compare with a recent
experiment which in our opinion has most likely already observed
this effect.

The problem is to some extent related to the one-hundred years old
Abraham-Minkowski debate on the correct electromagnetic
energy-momentum tensor in dielectric media.
 From a physical point of view the key issue
 is that one is dealing with a
{\it nonclosed} system, matter and field. Macroscopic of
phenomenological electromagnetic theory implying the use of a
permittivity $\varepsilon$ and permeability $\mu$ means that one
is dealing with a complicated interaction system involving
external fields, internal fields, and constituent molecules, by
using only simple material parameters. The solution to the problem
lies in extracting the energy-momentum form that leads to a
theoretical description of observable effects in a clean and
simple manner. For an overview of the Abraham-Minkowski debate,
cf.\ e.g.\ \cite{brevik79}. There exist at present a great number
of papers discussing the Abraham-Minkowski problem, for instance
the recent Ref.~\cite{hinds09} and the review \cite{pfeifer07}.

Assume now for definiteness that the fiber is vertically hanging,
and that a laser pulse is transmitted through it. The general
expression for the electromagnetic force density in the
medium, assumed hereafter nonmagnetic, is
 derived from
a given stress tensor $\sigma_{ik}$ as $f_i = \partial_k\sigma_{ik}$.
For the Abraham and Minkowski tensors this implies
(c.f.\ Refs.~\cite{brevik79,moller72} for details)
\begin{equation}
{\bf f}={\bf f}^\AM+\frac{n^2-1}{c^2}\frac{\partial}{\partial
t}\bf (\tE\times \tilde{\mathbf{H}}). \label{1}
\end{equation}
Here  ${\bf f}^\AM=-(\varepsilon_0/2)\tilde{E}^2{\bf \nabla}n^2$ is
nonvanishing in any region where $n$ varies, especially in the
surface regions. We use notation $\tE(\mathbf{r},t)=\mathrm{Re}\{\mathbf{E}(x)e^{i(\omega t-\beta z)}\}$ and skip the exponential factor in the following (c.f.~\cite{okamoto00} Ch.~1.3, for notational details). As this force is common for the Abraham and
Minkowski tensors, it may appropriately be denoted as ${\bf
f}^\AM$. The Abraham momentum density  ${\bf g}^A=(1/c^2){\bf
\tilde{E}\times \tilde{H}}$ occurs in the second term. We will henceforth ignore
it, due to the following two reasons: (i) for a stationary beam
the term fluctuates out when averaged over an optical period; (ii)
under perfectly axisymmetric conditions the force exerted on the
fiber during the transient entrance and exit periods has
necessarily to be vertical, thus being unable to initiate any
sideways motion.

In the following we investigate the effect of the force ${\bf f}^{AM}$ when the refractive index
contrast $\Delta n$ is assumed known from the mechanical
production process. Two simple planar models will be considered, in
order of increasing complexity. Finally we compare the theory with a recent experiment by She et al.~\cite{she08} and demonstrate how the deflection observed could very plausibly be a demonstration of the effect mentioned.

%%%%%%%%%%%%%%%%%%%%%%%%%%%%%%%%%%%%%%%%%%%%%%%%%%%%%%%%%%%%%%%
%%%%%%%%%%%%%%%%%%%%%% S E C T I O N %%%%%%%%%%%%%%%%%%%%%%%%%%
%%%%%%%%%%%%%%%%%%%%%%%%%%%%%%%%%%%%%%%%%%%%%%%%%%%%%%%%%%%%%%%
\section{Slab optical waveguide}

We will henceforth limit ourselves to planar geometries for the
fiber. This is mathematically simplifying, but the model is
nevertheless  expected to incorporate the essentials of the
imbalance effect. Probably  the simplest arrangement is to
consider a uniform  slab, infinite in the horizontal $y$
direction, having a finite width $2a$ in the other horizontal $x$
direction.

\begin{figure}[htb]
  \begin{center}
    \includegraphics[width=3.1 in]{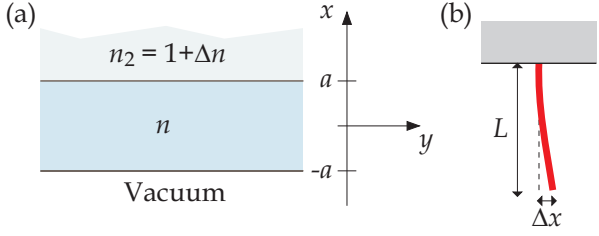}
    \caption{Left: The simplest model, a slab waveguide with a slight difference in refractive index above and below. Right: The geometry of
     a vertically hanging optical fiber subject to sideways motion.}
    \label{fig_simple}
  \end{center}
\end{figure}

The setup is sketched in Fig.~\ref{fig_simple}a; it is essentially the
same as Fig.~2.1 in Ref.~\cite{okamoto00}. The beam is propagating
into the plane, in the $z$ direction.
On the lower side of the slab ($x<-a$) we assume vacuum (or air),
with refractive index $n_0=1$. On the upper side ($x>a$), we
assume that there is a dilute medium, extending to $x=\infty$,
with refractive index $n_2=1+\Delta n$, $\Delta n \ll 1$. When
light propagates through the fiber, there will thus be an
imbalance in the surface force densities on the lower and upper
surfaces. Let use choose a TE mode (following the notation in
Ref.~\cite{okamoto00}). It means that the electric field has only
an $y$ component different from zero, ${\bf E}=(0,E_y,0)$. In the
dielectric boundary layers located around $x=-a$ and $x=a$, the
transverse force component is $f_x^\AM=-(\varepsilon_0/2)E_y^2\,
d (n^2)/dx$, yielding the respective surface force densities to be
\begin{align}
  \sigma_x(-a)=&-{\textstyle \frac{1}{4}}\varepsilon_0 E_y^2(-a)(n^2-1), \label{2}\\
  \sigma_x(a)=&-{\textstyle \frac{1}{4}}\varepsilon_0 E_y^2(a)(n_2^2-n^2).\label{3}
\end{align}
We have here taken into account that the longitudinal component
$E_y$ is continuous across the  surfaces, and that $\langle{\tilde{E}^2_y}\rangle = \frac{1}{2}E^2_y$ where $\langle~\rangle$ is average over oscillations in time and the $z$ direction.

The net transverse surface force density is
$\sigma_x=\sigma_x(-a)+\sigma_x(a)$. As $\Delta n$ is small, we
can make use of the expression for $E_y^2$ corresponding  to a
symmetric fiber, thus with the assumption $n_2=1$. We get
accordingly to first order in $\Delta n$
\begin{equation}
  \sigma_x=-{\textstyle \frac{1}{2}}\varepsilon_0 E_y^2(a)\cdot \Delta n, \label{4}
\end{equation}
where now $E_y^2(a)$ refers to the symmetric situation. We see
that for positive $\Delta n$, the surface force is directed
downwards in Fig.~\ref{fig_simple}a, i.e., in the negative $x$ direction. This is
as it should be, as surface forces are always directed towards the
optically thinner region at a dielectric surface.

The presence of $\sigma_x$ makes it possible to regard the fiber
as an elastic rod exposed to a constant transverse load. Let us
first, however, relate $\sigma_x$ to the total power $P$ in the
fiber. For the TE mode we have, using the same notation as in
Ref.~\cite{okamoto00},
\begin{equation}
  E_y=\left\{ \begin{array}{lll} A\cos (\kappa a-\phi)e^{-\xi(x-a)}, & x>a \\
  A\cos(\kappa x-\phi), & -a\leq x \leq a \\
  A\cos(\kappa a+\phi)e^{\xi (x+a)}, & x<-a.
  \end{array}
  \right. \label{5}
\end{equation}
Here $\kappa=\sqrt{n^2\omega^2/c^2-\beta^2}$ and
$\xi=\sqrt{\beta^2-\omega^2/c^2}$, where $\beta$ lying in the
interval $\omega/c \leq \beta \leq n\omega /c$ is the wave number
component in the $z$ direction. The corresponding nondimensional
transverse wave vectors are $u=\kappa a$, $w=\xi a$.

The
electromagnetic boundary conditions, requiring that $dE_y/dx$ be continuous, yield the following equations:
\begin{align}
  u=%&
  {\textstyle \frac{1}{2}}m\pi +\arctan \left(w/u\right),% \label{6}\\
  ~~~
  \phi=%&
  {\textstyle \frac{1}{2}}m\pi,% \quad m=0,1,2..., \label{7}
\end{align}
with $m=0,1,2...$, while $u$ and $w$ are related via
\begin{equation}
  u^2+w^2=\frac{\omega^2a^2}{c^2}(n^2-1). \label{8}
\end{equation}
From the above equations $\beta$ and $w$ can be calculated, and we
can find the relationship between $P$ and the constant $A$ in
Eq.~(\ref{5}) using formula  (2.34) in Ref.~\cite{okamoto00}:
\begin{equation}
  A^2=\frac{2\omega \mu_0 P}{\beta ab(1+1/w)}. \label{A}
\end{equation}
Recall that $P$ refers to the total power transmitted by  the
fiber. In the  planar model, we have let $b$ denote the fiber
width in the $y$ direction. The cross sectional area of the model
fiber is thus $2ab$, and the power per unit length in the $y$
direction is $P/b$. Edge effects because of the finite value of
$b$ are ignored. The transverse surface force density can now be
expressed as
\begin{equation}
  \sigma_x=-\frac{P}{abn_e c(1+1/w)}\frac{n^2-n_e^2}{n^2-1}\cdot \Delta n, \label{10}
\end{equation}
where we have defined
\be
  n_e = \beta c/\omega
\ee
and used that the continuity of $dE_y/dx$ across the interface at $x=a$ implies that
\be\label{coseq}
  \cos^2(u-\phi) = \cos^2(u+\phi) = \frac{\kappa^2}{\xi^2+\kappa^2} =\frac{n^2-n_e^2}{n^2-1}.
\ee

For practical purposes it may be convenient to express $\Delta n$
in Eq.~(\ref{10}) in terms of the corresponding increase $\Delta
\rho$ in material density. We can make use of the
Clausius-Mossotti relation, which is a good approximation at least
for nonpolar materials.  We then get
\begin{equation}
\Delta n^2=\frac{1}{3\rho}(n^2-1)(n^2+2)\cdot \Delta \rho.
\end{equation}

%HAVE REMOVED SYMBOL q.
Consider next the fiber as an elastic rod of rectangular cross
section, clamped at one end ($z=0$) and free at the other end
($z=L$). For convenience we let the $z$ axis be horizontal. We
choose $b=2a$, implying a square cross section of the fiber. The
transverse load per unit length in the longitudinal direction is
$\sigma_x b$ acting downwards in the negative $x$ direction.
The governing equation for the elastic deflection is (we ignore gravity)
$x''''(z)=\sigma_x b/(EI)$, where $E$ is Young's modulus and $I=b^4/12$ the
moment of inertia of the cross-sectional area  about its
centroidal axis \cite{landau70}. The solution of the governing
equation is
\begin{equation}
  x(z)=-\frac{\sigma_x b}{24EI}z^2(z^2-4Lz+6L^2). \label{11}
\end{equation}
The deflection at the tip, called simply $\Delta x$, is thus
\begin{equation}
  \Delta x=-\frac{\sigma_x bL^4}{8EI}. \label{12}
\end{equation}
We can now calculate the perturbation $\Delta n$ required to
produce a relative deflection $\Delta x/L$ at the tip:
\begin{equation}
  |\Delta n|=\frac{8EI}{L^3 P}n_e ac \left(1+\frac1{w}\right)\frac{n^2-1}{n^2-n_e^2}\cdot \frac{|\Delta x|}{L}. \label{13}
\end{equation}

In order to obtain an order of magnitude estimate for the required
difference in $n$ in order to yield a prescribed value for $\Delta
x$ we insert numerical values that are appropriate for a
low-intensity laser beam in a fiber (the numbers are typical, and the same as used in Ref.~\cite{she08}): $\lambda =
650$nm, $P=6.4$mW, $L=1.5$mm, $b=2a=450 $nm.

For the eigenvalues of the transverse wave number $\beta$ (or,
equivalently, $n_e$), corresponding to the guiding modes of the
planar fiber, we use again the solutions for the symmetrical
situations, since corrections to this only enter beyond leading
order in $\Delta n$. Following \cite{okamoto00}, the eigenvalue of
index $m=0,1,2,...$ solves \be
  f(b) = v\sqrt{1-b}-\arctan\sqrt{\frac{b}{1-b}}-\frac{m\pi}{2} = 0,
\ee
where $v = {\textstyle \frac{\omega a}{c}}\sqrt{n^2 - 1}$ and $b = (n_e^2 - 1)/(n^2-1)$. With the current numbers there are only two modes, $n_{e,0}=1.3635$ and $n_{e,1}=1.1072$.

As an example, assume now we desire a lateral displacement $\Delta x
= 15~\upmu$m. With the above numbers we find that the required $\Delta
n$ is only $5.5\cdot 10^{-4}$ for $m=0$ and $1.6\cdot 10^{-4}$ for
$m=1$. Minor mechanical defects from production could easily give
rise to changes in the reflective index of this magnitude. While
the planar model is likely to underestimate the required $\Delta
n$ slightly compared to a circular fiber, as an order of
magnitude estimate it demonstrates the feasibility of the scheme.

%%%%%%%%%%%%%%%%%%%%%%%%%%%%%%%%%%%%%%%%%%%%%%%%%%%%%%%%%%%%%%%
%%%%%%%%%%%%%%%%%%%%%% S E C T I O N %%%%%%%%%%%%%%%%%%%%%%%%%%
%%%%%%%%%%%%%%%%%%%%%%%%%%%%%%%%%%%%%%%%%%%%%%%%%%%%%%%%%%%%%%%
\section{Four-layered model}

\begin{figure}[tb]
  \begin{center}
    \includegraphics[width=2in]{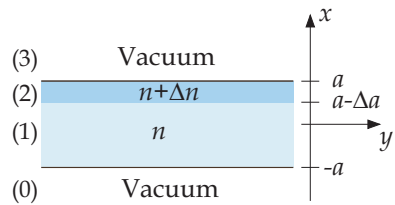}
    \caption{A two-layered model.}\label{fig_layer}
  \end{center}
\end{figure}

Regard now a slightly more realistic model where the slab has a layer of slightly higher refractive index on one side. The geometry is as
 considered in Fig.~\ref{fig_layer} where the slab, still of width $2a$ and refractive index $n$, has refractive index raised by $\Delta n$ in a layer of thickness $\Delta a$. Again we consider the TE mode, whose solution for the electric field component $E_y$ is written [in layers (3) to (0) from top to bottom]
\be
  E_y=A\left\{ \begin{array}{l} \cos (u-\Delta u-\phi)\cos(\Delta v-\psi)e^{-\xi
(x-a)}\\
\cos(u-\Delta u-\phi)\cos(\kappa' (x-a+\Delta a)-\psi),  \\
\cos(\kappa x-\phi)\cos\psi\\
\cos(u+\phi)\cos\psi \cdot e^{\xi (x+a)}
\end{array}
\right.
\ee
where $\Delta u= \kappa \Delta a, \Delta v = \kappa' \Delta a$, $\phi$ and $\psi$ are unknown phase angles and $\kappa^{\prime 2} = (n+\Delta n)^2k^2-\beta^2 \approx \kappa^2 + 2nk^2\Delta n$. The net surface force per unit area is now given by $\sigma_x = -(\varepsilon_0/2) \left.E_y^2(x)\right|_{x=-a}^{x=a}$. %, that is,
\begin{align}
  \sigma_x =& -{\textstyle \frac1{4}}\varepsilon_0 A^2[\cos^2 (u-\Delta u-\phi)\cos^2(\Delta v-\psi) \notag\\
  &-\cos^2(u+\phi)\cos^2\psi].
\end{align}
To leading order in $\Delta n$ we may again use the relation (\ref{A}) for $A$, and $\cos^2(u+\phi)$ is again given by equation (\ref{coseq}).

From continuity of $dE_y/dx$ at $x=a$ it follows that $\cos^2(\Delta v-\psi) = \kappa^{\prime 2}/(\xi^2+\kappa^{\prime 2})$, that is,
\[
  \cos^2(\Delta v-\psi)\approx \cos^2(u+\phi)\left(1-\frac{2n\Delta n}{n^2-1}\right)+\frac{2n\Delta n}{n^2-1}.
\]
Similarly, from the condition of continuity of $dE_y/dx$ at $x=a-\Delta a$ we derive that, to linear order in $\Delta n$ we have
\[
  \cos^2(u-\Delta u-\phi)\approx \cos^2\psi \left(1-\frac{2n\Delta n}{n^2-n_e^2}\sin^2\psi\right).
\]
Thus we find with some calculation that the surface force per unit area may be written as
\be
  \sigma_x \approx {\textstyle \frac1{2}}\varepsilon_0 A^2 \frac{n\Delta n}{n^2-1}\cos^2\psi \left(\frac{n^2-n_e^2}{n^2-1}-\cos^2 \psi\right).
\ee
We may express $\cos^2 \psi$ by means of the parameter $\Delta u$ by using the equation of continuity of $dE_y/dx$ at $x=a$, which may be written $\psi = -\arctan(\xi/\kappa) + \Delta v - m\pi$, $m=0,1,...$, to derive the relation, valid to leading order in $\Delta n$:
\be
  \cos^2\psi \approx \frac{n^2-n_e^2}{n^2-1}\left(\cos\Delta u + \frac{w}{u}\sin\Delta u\right)^2.
\ee
When $\Delta n$ is small we may use the same eigenvalues of $n_e$ as before.

\begin{figure}[tb]
  \begin{center}
    \includegraphics[width=2.6in]{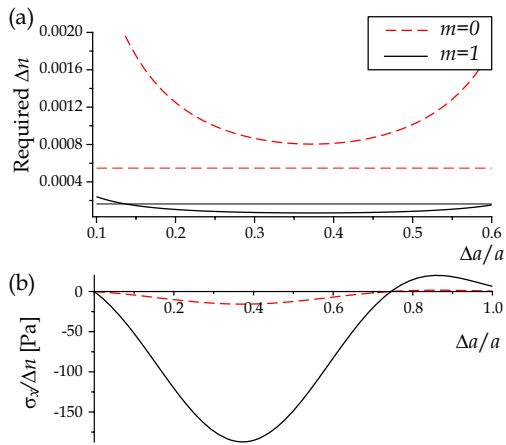}
    \caption{(a) Required $\Delta n$ as a function of $\Delta a/a$ to obtain $\Delta x=15\upmu$m with parameters as given in text. The thin straight lines show the corresponding values for the simple model of figure \ref{fig_simple}. (b) Surface force per unit area divided by $\Delta n$ as function of $\Delta a/a$.}\label{fig_dnbya}
  \end{center}
\end{figure}

Assuming once again that a displacement of $\Delta x=15\upmu$m is sought, figure \ref{fig_dnbya}a shows how the required $\Delta n$ changes with varying values of $\Delta a/a$. For the $m=0$ mode somewhat higher values of $\Delta n$ when $\Delta a/a$ is small. This is because $\sigma_x\propto \Delta n\Delta a$ as $\Delta a\to 0$. The required $\Delta n$ remains small, however, as long as the layer of slightly increased $n$ is non-vanishing. For $\Delta n > 1\%$ for example, $\Delta a/a$ must be smaller than about $0.035$ to achieve the observed deflection.

The surface force density $\sigma_x/\Delta n$ to leading order is
plotted as a function of $\Delta a/a$ in Fig.~\ref{fig_dnbya}b. As
in the simplest model of Fig.~\ref{fig_simple} the force is
directed downward when $\Delta a/a$ is
 smaller than approximately unity. In the present example it is reasonable to assume that it is a small number compared to
  $1$. We have performed the calculation also using different laser wavelengths, in which case the graph of
  Fig.~\ref{fig_dnbya}b looks qualitatively similar, but the abscissa is
rescaled. Since the point where the force density
   changes sign (about $\Delta a/a=0.75$ in Fig.~\ref{fig_dnbya}b) then moves along the abscissa, it is possible to change the sign of the force by changing the optical frequency, provided $\Delta a/a$ is chosen carefully.

\section{Comparison with an experiment. Conclusions and outlook}

The recent experiment of She {\it et al.} \cite{she08} appears to
be a natural example of application for the present theory. This
experiment tested precisely the sideways motion of a vertical
fiber when transmitted by a laser beam. Whereas the authors
actually related their observations to the Abraham-Minkowski
problem mentioned above, we do no think that such a conclusion is
right, and one of us has discussed this issue in more detail elsewhere
\cite{brevik09}. In the experiment \cite{she08} one side of the fiber was slightly harder than the
other from fabrication \cite{she09}, and it is natural
to suggest that this could have given rise to an appreciable $\Delta
n$.

Noting how even very small $\Delta n$ can cause deflections of the magnitude observed, and how the surface force depends on the geometry of the hardened layer in a nontrivial way,
it is much more likely in our opinion that all the
observed transverse deviations reported in \cite{she08} were due to the force ${\bf
f}^{\AM}$. As such the experiment is a striking demonstration of the theory presented herein.
%The values chosen for the laser power and the material parameters
%above, after Eq.~(\ref{13}), were actually taken from the same
%experiment, as these values are quite typical.
To ease comparison, our numerical examples above used data taken from the experiment \cite{she08}, including a deflection in the order of $15\upmu$m. A change in the refractive index on one side of the fiber in the order of less than $1\%$ is shown to be sufficient to produce deflections of such magnitude.
%The displacement
%of the end of the fiber was reported to be about $\Delta x \approx
%9\upmu$m about halfway through the pulse duration, but the
%equilibrium values are shown in Fig.~2 of \cite{she08} to be
%somewhat greater. We have therefore chosen the value  $\Delta x =
%15 \upmu$m in the theory above.

In conclusion, only a minute change in the refractive index, around $10^{-3}$, in a layer on on one side of an optical fiber of centimetre length and micron scale cross-section is sufficient to produce an appreciable
transverse deflection of the fibre. A tailored fiber of
this kind, where $\Delta n$ is prescribed accurately, should provide the possibility to bend fibers in a controlled and predictable
way. Because the lateral force acting on the fiber depends on the
laser frequency in a non-trivial way,
 it would moreover be possible to design a fiber which bends one way at one frequency, and changes direction of movement
  when a different frequency is used. The force can be used to carry a small load and one can envisage uses within
   micro- and nano robotics.

\bigskip
{\bf Acknowledgment}.  I. B. thanks Weilong She for
correspondence.

\end{document}